\documentclass[twocolumn,preprintnumbers,amsmath,amssymb,aps,prb,reprint]{revtex4}
\usepackage{graphicx}
\usepackage{color}

\begin{document}

\title{
Laning and Clustering Transitions in Driven Binary Active Matter Systems
} 
\author{
  C. Reichhardt$^{1}$,
  J. Thibault$^{1,2}$,
  S. Papanikolaou$^{2,3}$, and
  C.J.O. Reichhardt$^{1}$
}

\affiliation{
$^1$ Theoretical Division and Center for Nonlinear Studies,
Los Alamos National Laboratory, Los Alamos, New Mexico 87545, USA\\ 
$^2$ Department of Mechanical and Aerospace Engineering, Western Virginia University, Morgantown,
West Virginia 26506, USA\\ 
$^3$ Department of Physics, Western Virginia University, Morgantown, West Virginia 26506, USA\\
} 

\date{\today}
\begin{abstract}
It is well known that a binary system of non-active disks that experience driving in opposite directions exhibits jammed, phase separated, disordered, and laning states. In active matter systems, such as a crowd of pedestrians, driving in opposite directions is common and relevant, especially in conditions which are characterized by high pedestrian density and emergency. In such cases, the transition from laning to disordered states may be associated with the onset of a panic state.
   We simulate a laning system containing active disks that obey
  run-and-tumble dynamics, and we measure
  the drift mobility and structure 
  as a function of run length,  disk density, and drift force. 
  The activity of each disk can be quantified based on the correlation timescale of the velocity vector.
  We find that in some cases,  increasing the activity
  can increase the system mobility  by breaking up jammed configurations;
  however, an activity level that is too high
  can reduce the mobility by increasing the probability of disk-disk collisions.
  In the laning state, the increase of activity induces
  a sharp transition to
  a disordered strongly fluctuating state with reduced mobility.
  We identify a novel drive-induced clustered laning state
  that remains stable even at densities
  below the activity-induced clustering transition
  of the undriven system.
  We map out the dynamic phase diagrams
  highlighting transitions between the different phases as
  a function of activity, drive, and density.  
  \end{abstract}
\maketitle

\section{Introduction}

A binary assembly of interacting particles
that couple with opposite sign to an external drive such that the two
particle species move in opposite directions
has been
shown to exhibit a rich variety of dynamical behaviors \cite{1,2,3},
the most striking of which is
a transition to a laning
state
in which high mobility is achieved through organization of the particles
into noncolliding chains
\cite{3,4,5,6,7,8}.
Such systems have been experimentally realized
using certain types of colloidal particles \cite{9,10,11} 
or dusty plasmas \cite{12,13},
and have been used as a model for
motion in social systems
ranging from pedestrian flow 
\cite{2,14} to insect movement \cite{15}.
A variety of non-laning states can appear in these systems, including
jammed states where the
particles block each other's motion
\cite{16,17,18,19},
pattern forming states \cite{19,20,21,22,23,24,25},
and fully phase separated states \cite{17,18,19}. 
The laning transition has many similarities
to the phase separating patterns
observed in related driven binary systems,
indicating that formation of such patterns is a general
phenomenon occurring in many nonequilibrium systems 
\cite{26,27,28,29}.         
In a recent study of nonactive binary disks driven in opposite directions,
a comparison of the velocity force curves with 
those found in systems that exhibit depinning behavior revealed four
dynamic phases:
a jammed state, a fully phase separated high mobility state, a lower mobility 
disordered fluctuating state, and a
laning state \cite{19}.
The transitions between these phases as a function
of increasing drift force appear as jumps or features in the velocity force
curves coinciding with changes in the structural order of the system \cite{19}.  

Active matter, consisting of particles that can propel themselves independently of
externally applied forces, is an inherently nonequilibrium system commonly
modeled using either
driven diffusive or run-and-tumble dynamics \cite{30,31}.
For large enough activity, such systems are known to undergo
a transition from a uniform fluid state to a phase separated or clustered state 
\cite{32,33,34,35,36,37,38}.  The onset of clustering or swarming can
strongly affect the overall mobility of the particles
when obstacles or pinning are present \cite{39,40,41,42,43,44}. 
In studies of active matter moving under a drift force through obstacles,
the mobility is maximized at an optimal run length
since small levels of activity can break apart the clogging or jamming induced by
the quenched disorder, but high levels of activity generate self-induced
clustering that reduces the mobility \cite{44,45}.

In this work we examine
a binary system of oppositely driven active run-and-tumble particles.
In the absence of activity such a system is known to exhibit lane formation,
but we find that when activity is included, several new dynamic phases appear.
Adding activity to the non-active jammed state can break apart
the jammed structures and restore the mobility to finite values,
while when the non-active phase separated state is made active,
the system
becomes susceptible to undergoing jamming or clogging through a
freezing-by-heating effect \cite{46}.
High levels of activity generally decrease the mobility by producing
a disordered partially clustered fluctuating state.
The mobility of the non-active disordered state decreases when
activity is added,
while the non-active laning states undergo a sharp
transition as the activity is increased from low-collision, high mobility lanes to a
low mobility disordered state with
frequent particle collisions.
At high drives and large activity, we find a new phase
that we term a laning cluster phase in which
dense clusters appear that are phase separated
into the two different oppositely driven species.
The laning cluster phase
is stable down to particle densities well below the onset of
activity-induced clustering in an undriven system.
Transitions among these different phases can be identified through
changes in mobility, changes in the particle structure, or
changes in the frequency of particle-particle collisions, and we use
these changes to
map the dynamic phases as a function of 
external drift force, density, and activity.
We draw analogies between the
sharp transition we observe from the high mobility laning state to the low mobility 
disordered fluctuating state
and panic transitions in which a high mobility state of
pedestrian flow can change into a low mobility panic state
in which continuous collisions between pedestrians occur.

We note that previous work on oppositely driven
active matter particles by Bain and Bartolo \cite{47} focused
on the nature of the critical behavior at the transition between a
fully phase separated state and a disordered mixed phase,
rather than the mobility that we consider.
Reference \cite{47} also uses
a flocking or Vicsek model, which is distinct from the
run-and-tumble or driven diffusive
active matter systems that are the focus of our work.

\section{Simulation and System}
We consider a two-dimensional system of size $L \times L$
with periodic boundary conditions in the $x$ and $y$ directions
containing $N$ particles of radius $R_{d}$.
We take $L=36$ and $R_d=0.5$.
The interaction between particles $i$ and $j$ 
has the repulsive harmonic form
${\bf F}^{ij}_{pp} = k(r_{ij} -2R_{d})\Theta(r_{ij} -2R_{d}){\bf \hat r}_{ij}$,
where $r_{ij} = |{\bf r}_{i} - {\bf r}_{j}|$, 
$ {\bf \hat r}_{ij} = ({\bf r}_{i} - {\bf r}_{j})/r_{ij}$,
and $\Theta$ is the Heaviside step function. 
We set the spring stiffness $k = 50$
large enough that there is less than a one percent overlap between the
particles, placing us in the
hard disk limit as confirmed in previous works \cite{11,12,37}.
The area coverage of the 
particles is $\phi = N \pi R^{2}_{d}/L^2$, and
a triangular solid forms for $\phi = 0.9$ \cite{37}.
The particles are initialized in non-overlapping randomly chosen
locations and are coupled to an external dc drift force
${\bf F}_d=\sigma_i F_d {\bf \hat x}$,
where $\sigma_i=+1$ for half of the particles, chosen at random, and
$\sigma_i=-1$ for the remaining half of the particles.
The dynamics of particle $i$
are determined by the following overdamped equation of motion:
\begin{equation}
\eta \frac{d{\bf r}_i}{dt} = \sum_{j \neq i}^{N}{\bf F}_{pp}^{ij} + {\bf F}_{d} + {\bf F}_{m}^i 
\end{equation}
Each particle experiences a motor force ${\bf F}_{m}^i=F_m{\bf \hat{\xi}}$
which propels the particle in a randomly chosen direction ${\bf \hat{\xi}}$
for a fixed run time $\tau$.
At the end of each run time, the particle tumbles instantaneously by selecting
a new randomly chosen direction for the next run time.
The amplitude of the motor force is $F_{m} = 1.0$
and the simulation time step is $\delta t = 0.002$, so in the
absence of other forces a particle moves a distance
called the run length
$l_{r} = F_{m}\delta t \tau$ during each run time.
To increase the activity of the particles, we increase $\tau$ so that the
correlation time of the self-driven motion becomes larger.
After applying the dc drive, we measure
the time average of the velocity per particle
for only the $\sigma_i=+1$ particles in the $+x$ dc drift direction,
$\langle V\rangle = (2/N)\sum^{N}_{i =1}\delta(\sigma_i-1)({\bf v}_{i}\cdot {\bf \hat x})$,
where ${\bf v}_i$ is the velocity of particle $i$.
The corresponding average velocity in the drift direction curve for the
$\sigma_i=-1$ particles is identical to $\langle V\rangle$ due to symmetry.
 We wait a minimum of
 $10^7$ simulation time steps before taking the measurement
 to ensure that the system has reached a steady state.

 \section{Laning and Clustering at Low Densities}

\begin{figure}
\includegraphics[width=\columnwidth]{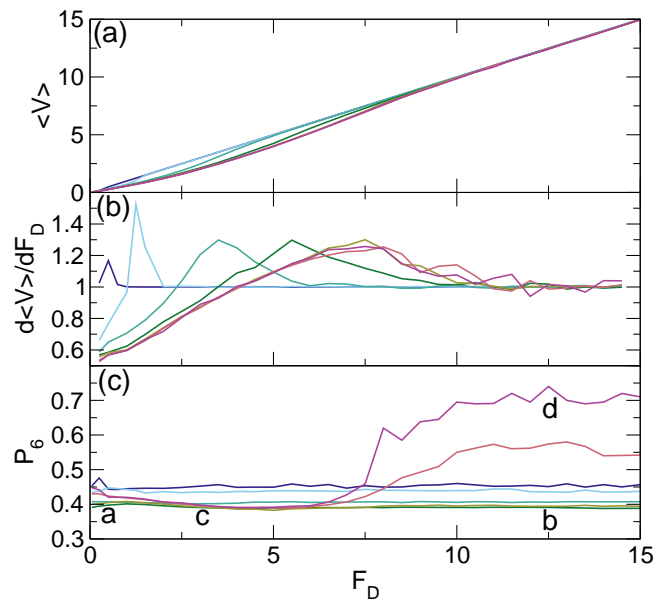}
\caption{ (a) The average velocity per particle $\langle V\rangle$ in the
  dc drift direction
  for the $\sigma_i=+1$ particles
  vs $F_{D}$ 
  in a sample with particle density $\phi=0.424$.
  The run-and-tumble particles have run time
$\tau = 10$ (dark blue), $30$ (light blue), $150$ (light green), $500$ (dark green),
$2000$ (gold), $2 \times 10^4$ (pink) and $3.2 \times 10^5$ (magenta). 
  (b) The corresponding $d\langle V\rangle/dF_{D}$
  vs $F_D$ curves showing a peak that shifts to higher values of $F_D$ as $\tau$ increases.
  (c) The corresponding fraction of sixfold-coordinated particles
  $P_{6}$ vs $F_{D}$. The $\tau=2 \times 10^4$ and $\tau=3.2 \times 10^5$
  curves show a transition to a state with
  high triangular ordering,
  indicative of clustering.    
  The letters {\bf a}, {\bf b}, {\bf c}, and {\bf d}
  mark the values of $F_{D}$ at which
  the images in Fig.~\ref{fig:2} were obtained.
}
\label{fig:1}
\end{figure}
 
\begin{figure}
\includegraphics[width=\columnwidth]{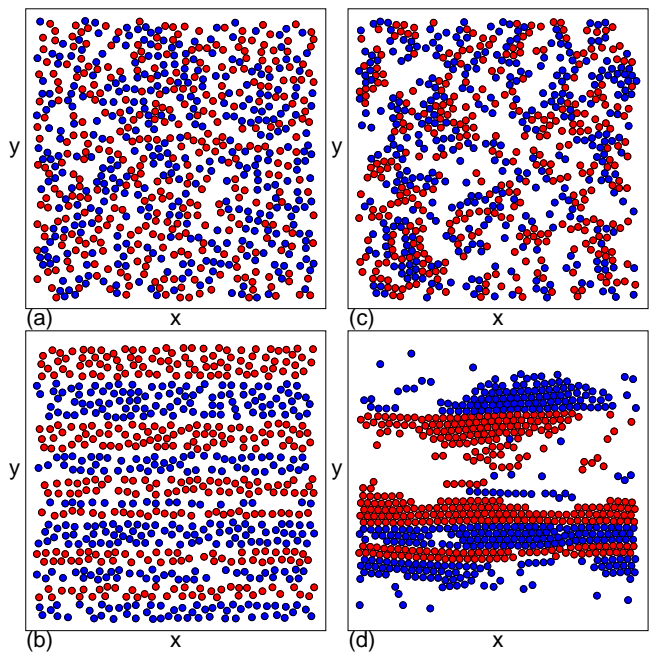}
\caption{ Instantaneous positions of the $\sigma_i=+1$ (blue) and $\sigma_i=-1$ (red)
  run-and-tumble particles subjected to a drift force $F_D$.
  (a) A phase III mixed liquid state at $\tau = 500$ and $F_{D} = 0.5$.
  (b) A phase IV laning state at $\tau = 500$ and $F_{D} = 12.5$.
  (c) A phase III mixed state with local clustering at
  $\tau = 3.2 \times 10^5$ and $F_{D} = 3.0$.
  (d) A laning cluster phase V at $\tau = 3.2 \times 10^5$ and $F_{D} = 12.5$.
  The particles in the clusters have a significant amount of triangular ordering,
  producing an increase in $P_6$ in phase V in Fig.~\ref{fig:1}(c).
}
\label{fig:2}
\end{figure}

Previous work on non-active laning particles identified
four dynamic phases: a jammed state (phase I), a fully phase separated state (phase II),
a mixed or disordered state (phase III), and a laning state (phase IV) \cite{19}.
For particle densities $\phi < 0.55$, the system is always in a laning state,
while for $\phi \geq 0.55$, the other three phases appear as well.
For the active particles, we adopt the same nomenclature for phases I to IV,
and define the low density regime as $\phi < 0.55$.
In Fig.~\ref{fig:1}(a) we plot $\langle V\rangle$
versus $F_{D}$ for a sample with $\phi=0.424$ for run lengths ranging from
$\tau=10$ to $\tau=3.2 \times 10^5$.
All of the velocity-force curves have nonlinear behavior at low drives that
transitions to a linear response at higher drives, as indicated by the
peak in the $d \langle V\rangle/dF_D$ versus $F_D$ curves in
Fig.~\ref{fig:1}(b).
The nonlinear behavior extends up to higher values of $F_D$ as $\tau$ increases.
For $\tau < 1.5 \times 10^4$, the peak in
$d\langle V\rangle/dF_{D}$ coincides with
the transition from disordered phase III flow
to laning phase IV flow.
Thus, as the activity is increased by raising $\tau$,
higher drift forces $F_D$ must be applied in order to induce
lane formation.
In Fig.~\ref{fig:2}(a) we illustrate
the particle positions
at $\tau = 500$ and $F_{D} = 0.5$ in the phase III disordered or mixed liquid state, 
while in Fig.~\ref{fig:2}(b) we show the same system
in phase IV at  $F_{D}= 12.5$ where the
particles form stable lanes.

In Fig.~\ref{fig:1}(c) we plot the fraction of
sixfold-coordinated particles $P_6$ versus $F_D$.  Here,
$P_6=N^{-1}\sum_{i=1}^{N}\delta(z_i-6)$ where the coordination number $z_i$ of particle
$i$ is obtained from a Voronoi construction.
For $\tau < 1 \times 10^4$, there is 
no clear jump in $P_6$ at the transition from
phase III to phase IV since,
as shown in Fig.~\ref{fig:2}(b),
the flowing lanes have no crystalline ordering.
For $\tau > 1.5 \times 10^4$,
phase IV is replaced by a new phase V, as indicated
by the increase in $P_{6}$ at large $F_D$ for the
$\tau = 2 \times 10^4$ and $\tau = 3.2 \times 10^5$ curves.
Phase V is what we term a clustered laning state, as illustrated
in Fig.~\ref{fig:2}(d) at
$\tau =3.2 \times 10^5$ and $F_{D} = 12.5$.
Here the particles form clusters similar to the activity-induced
clusters that appear in an undriven active matter system \cite{32,33,34,35,36,37,38},
but within each cluster, phase segregation into the two oppositely moving
particle species occurs in order to eliminate particle-particle collisions.
Triangular ordering of the particles emerges within the denser clusters,
leading to the increase in $P_{6}$ 
at the onset of phase V.
For the same large $\tau = 3.2 \times 10^5$ at a lower drive of $F_D=3.0$,
a phase III disordered mixed phase occurs as illustrated in
Fig.~\ref{fig:2}(c), where a small amount of clustering is visible due to
the high activity level.

\begin{figure}
\includegraphics[width=\columnwidth]{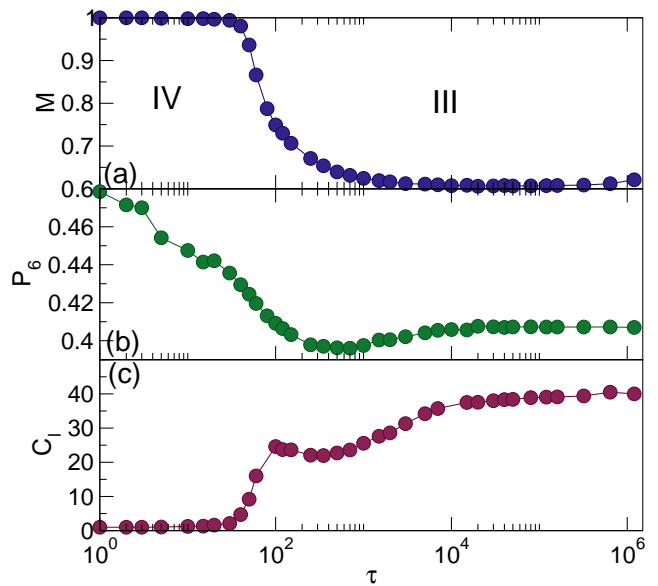}
\caption {(a) Mobility $M$ vs $\tau$ for a system with $\phi =0.424$ and $F_{D} = 2.0$.  
  (b) The corresponding $P_{6}$ vs $\tau$.
  (c) The corresponding average largest cluster size $C_l$ vs $\tau$.
  The IV-III transition that occurs with increasing $\tau$ is associated
  with a drop in $M$, a decrease in $P_6$, and
  an increase in $C_{l}$.
}
\label{fig:3}
\end{figure}

We define the mobility
$M=\langle V\rangle/V_0$
as the average
particle velocity divided by the expected free flow velocity $V_0=F_D/\eta$
of an individual particle in the absence of particle-particle interactions.
In Fig.~\ref{fig:3}(a) we plot $M$
versus $\tau$ for a system
with $\phi = 0.424$ at $F_{D} = 2.0$, where $V_{0} = 2.0$.
For $\tau < 100$ the system forms a
phase IV laning state similar to that
illustrated in Fig.~\ref{fig:2}(b),
and as $\tau$ increases, a transition to phase III occurs
that is accompanied by a sharp decrease in the mobility
from $M=1.0$ to $M=0.62$.
The corresponding $P_6$ versus $F_D$ curve appears
in Fig.~\ref{fig:3}(b), showing that
$P_{6}$ decreases with increasing $\tau$
but has
no sharp feature
at the IV-III transition.
In Fig.~\ref{fig:3}(d) we plot $C_l$, the average largest cluster size, versus $F_D$ for
the same system.  To measure $C_l$, we group the particles into clusters by
identifying all particles that are in direct contact with each other, determine the
number of particles $N_c^j$ in a given cluster $j$, and obtain
$C_l=\langle \max\{N_c^j\}_{i=1}^N\rangle$ where the average is taken over a series of
simulation time steps.
Larger values of $C_l$ indicate that particle-particle collisions are more frequent.
In steady state phase IV flow,
the particles only experience brief pairwise collisions,
so $C_{l} < 3$; additionally,
the mobility is close to $M=1$ since the
particles are undergoing nearly free flow.
At the IV-III transition, the particle collision frequency increases,
lowering the mobility, while
the cluster size increases, with $C_l$ reaching values of 30 or more.  
The IV-III transition that occurs when $\tau$ increases can be
regarded as analogous to a transition in a social system
from orderly laning flows of noncolliding pedestrians
to a panic state in which
pedestrians collide and impede each other's flow.
Here, the run time would correspond to an agitation level which,
above a certain threshold, destroys the orderly flow and produces
a low mobility collisional flow.

\begin{figure}
\includegraphics[width=\columnwidth]{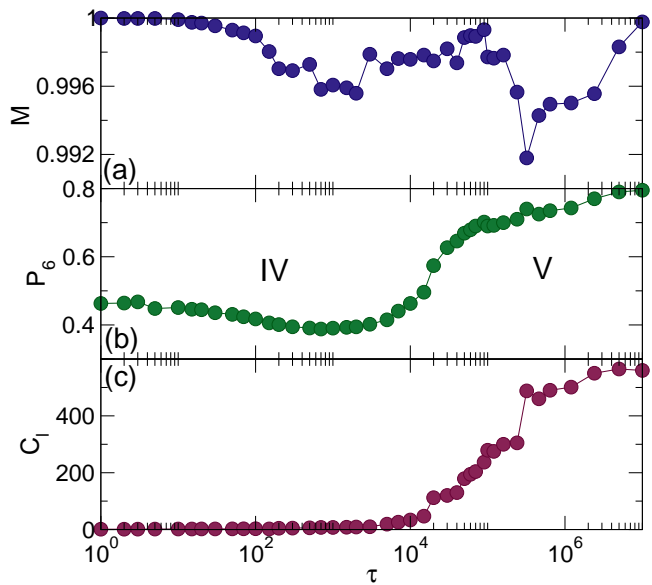}
\caption
{
(a) Mobility $M$ vs $\tau$ for a system with $\phi =0.424$ at $F_{D} = 12.5$.
  (b) The corresponding $P_{6}$ vs $\tau$.
  (c) The corresponding average largest cluster size $C_{l}$ vs $\tau$.
  At the IV-V transition,
  both $P_6$ and $C_l$ increase, but there is little change in $M$.
}
\label{fig:4}
\end{figure}

In Fig.~\ref{fig:4}(a--c) we plot $M$, $P_{6}$, and $C_{l}$ versus $\tau$
for the $\phi=0.424$ system from Fig.~\ref{fig:3} at a higher drive
of
$F_{D} = 12.5$, where very different behavior appears.
At low $\tau$ the system is initially in the phase IV laning state due to the
large drive,
and as $\tau$ increases, a transition occurs into the clustered laning phase V
illustrated in
Fig.~\ref{fig:2}(d),
rather than the disordered phase III flow that appears at lower $F_D$.
In phase IV, $C_{l}$ is low since particle collisions are rare, and
$P_6 \approx 0.5$ due to the one-dimensional liquid structure of the
flow.
At the transition
to phase V,
both $C_{l}$ and $P_{6}$ increase
to  $C_l \approx 500$ and $P_{6} \approx 0.8$, while there is very little
change in the mobility $M$.
Unlike the mixed flow found in phase III,
phase V is is mostly phase separated as shown in Fig.~\ref{fig:2}(d),
so the mobility is high even though $C_l$ is large, since the particle-particle
interactions
are dominated by static contacts within the moving clusters rather
than by collisional contacts between clusters moving in opposite directions.

\begin{figure}
\includegraphics[width=\columnwidth]{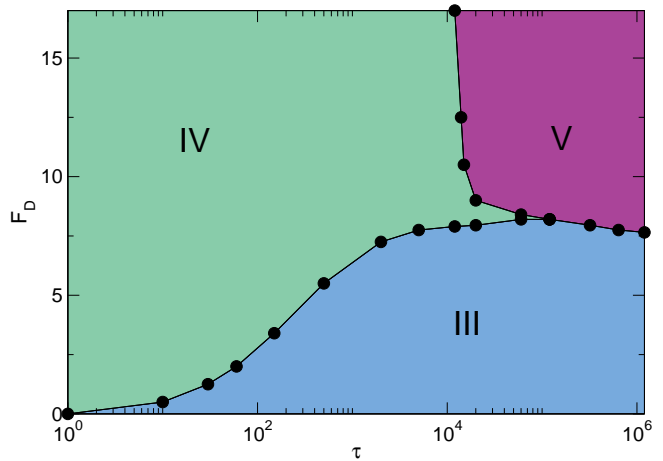}
\caption
{
  Dynamic phase diagram as a function of
  $F_{D}$ vs $\tau$ at $\phi=0.424$.  Phase III: disordered mixing flow state;
  phase IV: laning state; phase V: clustered laning state.
}
\label{fig:5}
\end{figure}

In Fig.~\ref{fig:5} we construct a dynamic phase diagram
as a function of $F_D$ versus $\tau$ for a system with $\phi=0.424$ highlighting
the regimes of phase III, IV, and V flow.
The transitions
between the phases
are identified based on changes in $M$, $P_{6}$ and $C_{l}$.
For $F_{D} < 8.0$, the system is
in phase IV at small $\tau$ and phase III at large $\tau$, as
illustrated in Fig.~\ref{fig:3}.
For
$0 < \tau < 1.5 \times 10^4$, the IV-III transition line shifts to
larger $F_D$ with increasing $\tau$.
For $\tau > 1.5 \times 10^5$ and $F_{D} > 8.0$, the system is
still in phase IV at small $\tau$ but is in phase V at large $\tau$, as
shown
in Fig.~\ref{fig:4}.
We note that at this particle density of $\phi=0.424$,
when $F_{D} = 0$
there is no activity-induced clustered state,
since as shown in previous studies of this model in a similar regime, such a state arises only
for
$\phi > 0.45$
\cite{19}. 
The results in Fig.~\ref{fig:5} indicate that driving can induce the formation
of a clustered state at large activity even at particle densities for which
activity alone cannot produce a clustered state.
This suggests that active non-clustering
fluid states
could transition to a clustered state
under application of a shear or other external driving.  

\section{Drive Induced Active Phase Separation}

\begin{figure}
\includegraphics[width=\columnwidth]{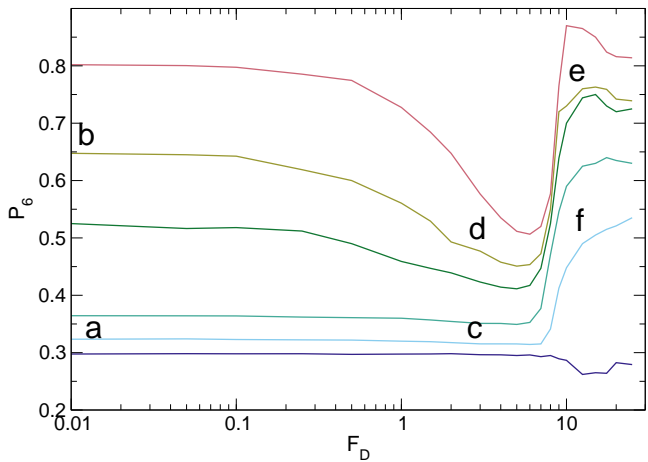}
\caption
    { $P_{6}$ vs $F_{D}$ at $\tau = 3.2 \times 10^5$ for
      $\phi = 0.06$, 0.182, 0.303, 0.48, 0.6, and $0.848$,
      from bottom to top. 
      The letters {\bf a} to {\bf f} indicate the points at which the images
      in Fig.~\ref{fig:7} were obtained.
}
\label{fig:6}
\end{figure}

\begin{figure}
\includegraphics[width=\columnwidth]{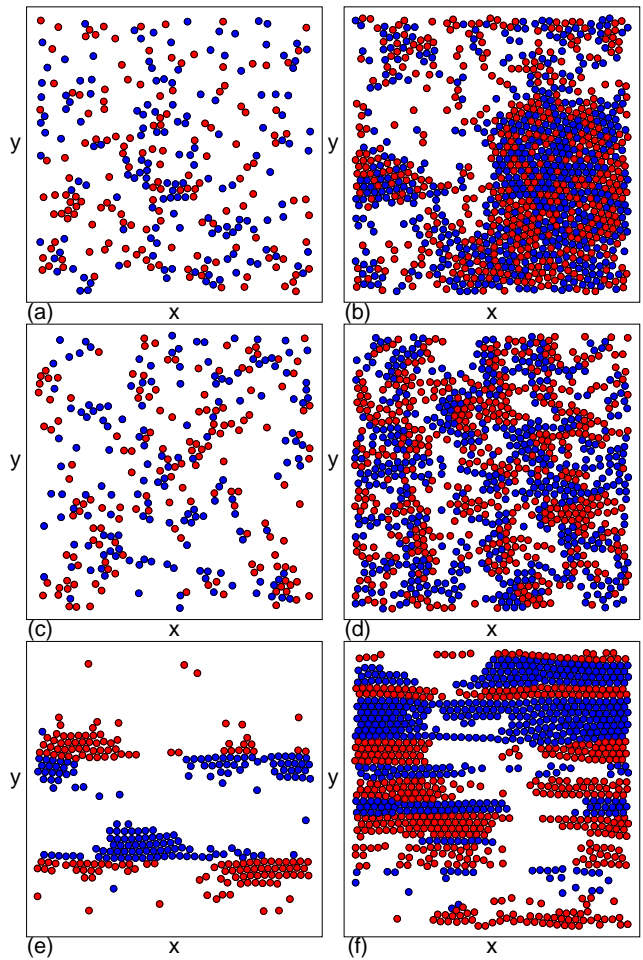}
\caption
    { Instantaneous positions of the $\sigma_i=+1$ (blue) and
      $\sigma_i=-1$ (red) run-and-tumble particles 
      from the system in Fig.~\ref{fig:6} at $\tau = 3.2 \times 10^5$.
      (a) Phase III at $\phi = 0.182$ and $F_{D} = 0.01$.
      (b) At $\phi = 0.6$ and $F_{D} = 0.01$,
      a cluster state forms with no separation of the different species.
      We call this phase CL.
      (c) Phase III at $\phi = 0.182$ and $F_{D} = 3.0$.
      (d) Phase III with weak clustering at
$\phi = 0.6$ and $F_{D} = 3.0$.
      (e) Phase V, the laning cluster state,
      at $\phi = 0.182$ and $F_{D} = 12.5$.
      (f)  Phase V at $\phi = 0.6$ at $F_{D} = 12.5$.
}
\label{fig:7}
\end{figure}

\begin{figure}
\includegraphics[width=\columnwidth]{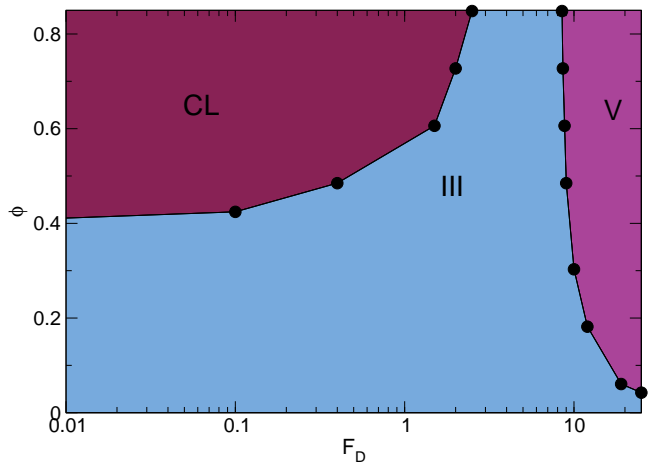}
\caption
    {Dynamic phase diagram as a function of $\phi$ vs $F_{D}$
      for the system in Figs.~\ref{fig:6} and \ref{fig:7} at $\tau = 3.2 \times 10^5$. 
      At small $F_{D}$, there is a transition from phase III to a cluster state CL
      with increasing $\phi$, while large drives can produce
      phase V. 
}
\label{fig:8}
\end{figure}

We next study the evolution of phases IV and V in greater detail
over a range of particle densities and external drives.
In Fig.~\ref{fig:6} we  plot $P_{6}$ versus $F_{D}$ at
$\tau = 3.2 \times 10^5$ for particle densities ranging from
$\phi = 0.06$
to
$\phi=0.848$.
At the lowest values of $F_{D}$, when $\phi<0.475$,
$P_6<0.45$ and the system is
always in a disordered state as
illustrated in Fig.~\ref{fig:7}(a) at  $\phi = 0.182$ and $F_{D} = 0.01$.
When $\phi > 0.4$, there is a transition to a cluster state
in the absence of drive, and this cluster state, which
we term phase CL, persists at low
drives,
as shown in Fig.~\ref{fig:7}(b) for $\phi = 0.6$ and $F_{d} = 0.01$.
Here a dense solid-like region with a significant amount of
triangular ordering is surrounded by a low-density liquid.
The cluster state has a density phase separation into high and
low density regions; however, there is no segregation of the two
particle species, which distinguishes phase CL from the laning cluster
phase V.
At intermediate values of $F_{D}$, the disordered
flow phase III appears as shown in
Fig.~\ref{fig:7}(c) for $\phi = 0.182$ and $F_{D} = 3.0$.
The larger $F_D$ value tears apart the cluster state for $\phi>0.4$,
producing in its place a disordered phase III flow with some residual
clustering, as illustrated in
Fig.~\ref{fig:7}(d)
at $\phi = 0.6$ and $F_{D} = 3.0$.
In Fig.~\ref{fig:7}(e,f) we show the $F_D=12.5$ states at
$\phi = 0.182$ and $\phi=0.6$, respectively.
In both cases
a laning cluster phase V appears,
producing
the higher values of $P_{6}$ found in Fig.~\ref{fig:6}.
Phase V persists
all the way down to $\phi = 0.06$
for this high drive;
however, at the smaller values of $\phi$
the phase separated regions become more one-dimensional in
nature, so $P_6$ remains low due to the smaller coordination number
of the particles in these chain-like structures.
Based on the features in Fig.~\ref{fig:7}
along with additional simulation data,
we construct a dynamic phase diagram as a function of
$\phi$ versus $F_D$ for $\tau=3.2 \times 10^5$ as shown
in Fig.~\ref{fig:8}.
This result suggests that the introduction of
shearing or driving can break up the clusters that form due to
activity-induced density segregation; however,
at large enough shearing, a new type of clustering instability can arise.

\begin{figure}
\includegraphics[width=\columnwidth]{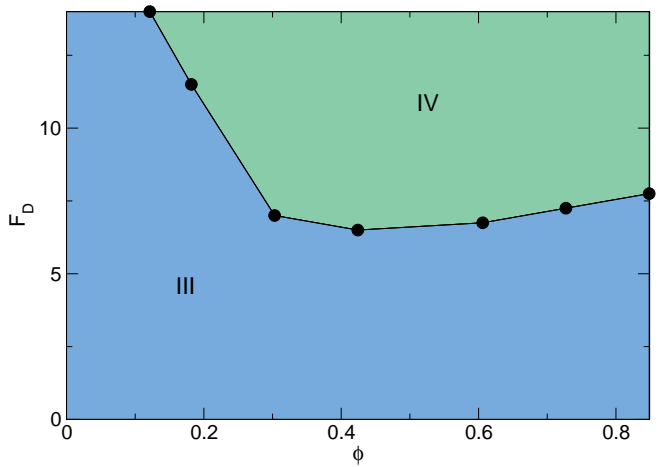}
\caption
{ Dynamic phase diagram as a function of $\phi$ vs $F_{D}$ at $\tau = 500$
showing phases III and IV. 
}
\label{fig:9}
\end{figure}

\begin{figure}
\includegraphics[width=\columnwidth]{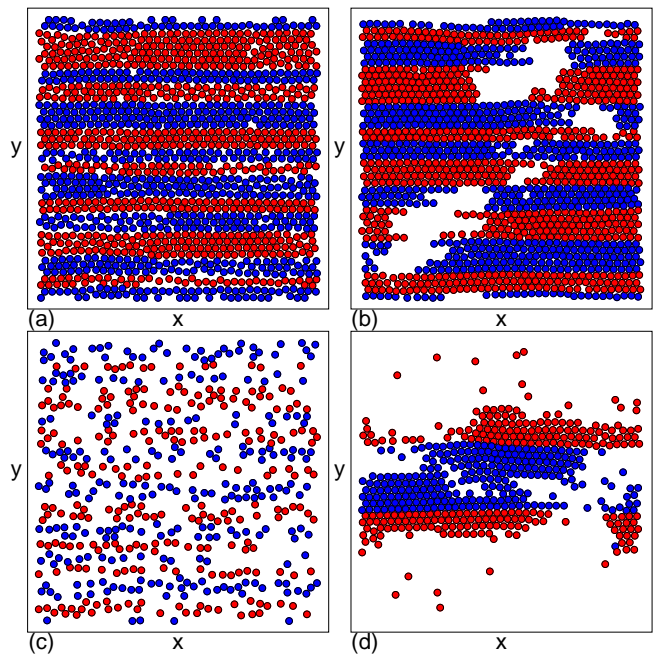}
\caption{ A comparison of the instantaneous
  particle configurations of the $\sigma_i=+1$ (blue) and
  $\sigma_i=-1$ (red) run-and-tumble particles for the systems
  in Figs.~\ref{fig:8} and \ref{fig:9} at $F_D=12.5$.
  (a,b) $\phi=0.848$:
  (a) The laning phase IV for the system in Fig.~\ref{fig:9}
  with $\tau = 500$.
  (b) The laning clustered phase V for the system in Fig.~\ref{fig:8}
  with $\tau = 3.2 \times 10^5$.
  (c,d) $\phi=0.303$:
  (c) A laning phase IV without triangular ordering for the system in
  Fig.~\ref{fig:9}
  with
  $\tau = 500$.
  (d) The laning clustered phase V for the system in Fig.~\ref{fig:8}
  with $\tau = 3.2 \times 10^5$.
}
\label{fig:10}
\end{figure}

In Fig.~\ref{fig:9} we plot a dynamic phase diagram as a function of
$F_{D}$ vs $\phi$ for a small run time of $\tau = 500$. 
In this case, neither the phase CL nor phase V appear.
In Fig.~\ref{fig:10}(a)
we show the
instantaneous particle configuration in the laning phase IV
for the system in Fig.~\ref{fig:9}
at  $F_{D} = 12.5$, $\phi=0.848$, and $\tau=500$,
while in Fig.~\ref{fig:10}(b) at
the same values of $F_D$ and $\phi$ but with $\tau=3.2 \times 10^5$
as in Fig.~\ref{fig:8},
the system
forms a laning clustered state containing low density regions.
At $F_D=12.5$ and $\phi=0.303$,
Fig.~\ref{fig:10}(c) shows
that the $\tau = 500$ system from Fig.~\ref{fig:9}
enters a phase IV flow with no triangular ordering,
while in Fig.~\ref{fig:10}(d),
the $\tau=3.2 \times 10^5$ system from Fig.~\ref{fig:8}
forms the laning clustered phase V.

This system
could also serve as a soft matter realization
of certain types of social dynamics such as pedestrian flows, and could
be used to study the transition
from orderly laning flow to disordered or panic motion. In this context, if Phase III is identified as disorderly pedestrian flow and Phase IV as orderly flow, then our main conclusion would be that the activity timescale strongly influences the force required to approach the transition, as in Fig.~\ref{fig:5},
while further increasing an already large pedestrian density may not be
such a significant factor, as in Fig.~\ref{fig:9}.

\section{Dense Phase}

\begin{figure}
\includegraphics[width=\columnwidth]{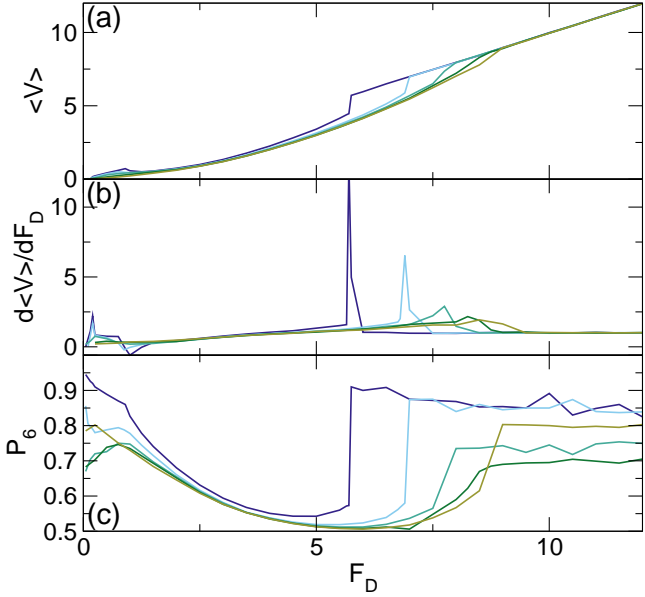}
\caption{ A system at $\phi = 0.848$ with
  $\tau = 10$ (dark blue), $150$ (light blue), $500$ (light green), $2000$ (dark green), 
  and $2 \times 10^4$ (gold).
  (a) $\langle V\rangle$ vs $F_{D}$.
  (b) $d\langle V\rangle/dF_{D}$ vs $F_{D}$.
  (c) $P_{6}$ vs $F_{D}$. 
  For $\tau < 500$ we observe phase I (jammed), II (phase separated), III (disordered
  mixed flow), and IV (laning flow).
  Transitions between these phases appear as features in
  $d\langle V\rangle/dF_D$:
  an initial spike near $F_D=0.15$ is the I-II transition,
  a negative region near $F_D=1.0$ is the II-III transition, and
  the large spike that appears for $F_D>5.0$ is the III-IV transition.
  For $\tau > 1.5 \times 10^4$, the III-IV transition
  is replaced by a III-V transition.
}
\label{fig:11}
\end{figure}

\begin{figure}
\includegraphics[width=\columnwidth]{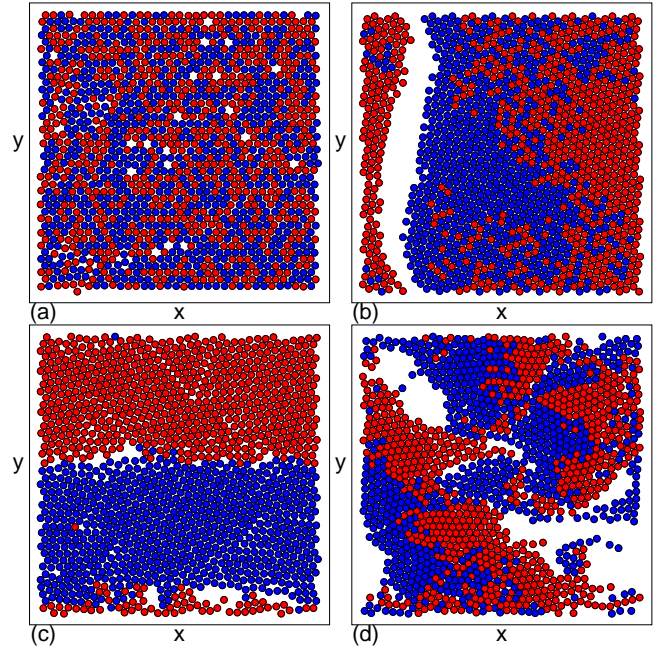}
\caption{Instantaneous positions of the $\sigma_i=+1$ (blue) and
  $\sigma_i=-1$ (red) run-and-tumble particles in
  the system from Fig.~\ref{fig:11} with $\phi = 0.848$ and $\tau =10$.
  (a) The jammed phase I at $F_{D} = 0.01$.
  (b) The jammed phase I at $F_{D} = 0.15$.
  (c) The phase separated state (phase II) at $F_{D} = 0.5$.
  (d) The disordered mixed flow phase III at $F_{D} = 1.5$. }
\label{fig:12}
\end{figure}

We next consider the role of activity in the dense phase
with $\phi = 0.848$.
Here, when $F_D<1.0$, we find two additional phases: a
jammed state (phase I) and a phase separated state (phase II).
In Fig.~\ref{fig:11}(a) we plot representative
$\langle V\rangle$ versus $F_{D}$ curves
for run times ranging from $\tau=10$ to $\tau=2 \times 10^4$.
Figure~\ref{fig:11}(b) shows the corresponding
$d\langle V\rangle/dF_{D}$ versus $F_{D}$ curves and in Fig.~\ref{fig:11}(c) 
we plot $P_{6}$ versus $F_{D}$.
For $\tau < 500$ we find the jammed phase I in which
$\langle V\rangle =0$ and
$d\langle V\rangle/dF_{D} = 0$.
The particle configurations in the two variations of the
jammed state are illustrated in
Fig.~\ref{fig:12}(a,b) for $\tau=10$  at $F_D=0.01$ and $F_D=0.15$, respectively.
When $F_{D} < 0.05$, the system remains in its initially deposited configuration
and only small rearrangements occur before the particles settle into a
motionless jammed state,
while for $F_{D} > 0.05$, the system undergoes transient large-scale rearrangements
before organizing into a jammed state of the type illustrated in Fig.~\ref{fig:12}(b).
Here there is both a density phase separation into high and zero density regions
as well as a species phase separation, with the
$\sigma_i=-1$ particles preferentially sitting to the left of the $\sigma_i=+1$ particles
and blocking their motion.
In phase II, the phase separated state illustrated in Fig.~\ref{fig:12}(c) for
$F_D=0.5$, each particle species forms a mostly triangular solid,
giving a large value of $P_{6}$.
The I-II transition is associated with
a spike in the $d\langle V\rangle/dF_{D}$ curves near $F_{D} = 0.15$. 
Within phase II, the phase separation allows
the particles to move without collisions,
so individual particles move at nearly the free flow velocity $V_0$ and the mobility
$M \approx 1$.
As $F_{D}$ increases, a II-III transition occurs.
We illustrate the disordered mixed flow phase III
in Fig.~\ref{fig:12}(d) for $\tau = 10$ and  $F_{D} = 1.5$.
The particles are in a fluctuating state and undergo numerous collisions,
reducing the mobility.
Just above the transition into phase III, some clustering of the particles persists,
as shown in Fig.~\ref{fig:12}(d).  As $F_D$ increases, the size of these
clusters drops, causing $P_6$ to decline.
The II-III transition
is associated with a drop 
in $\langle V\rangle$ and $P_6$ along with
negative values of $d\langle V\rangle/dF_{D}$,
indicative of negative differential conductivity,
For $F_{D} > 5.0$ and $\tau < 1.5 \times 10^4$, the system transitions
from phase III to the laning cluster phase IV as shown previously,
and this transition corresponds
with upward jumps in $\langle V\rangle$ and $P_6$ and a large positive spike 
in $d\langle V\rangle/dF_{D}$.
For $\tau > 500$, phases I and II disappear,
as indicated by the
loss of the spikes in $d\langle V\rangle/dF_{D}$
and the reduced value of $P_{6}$ at small values of $F_{D}$.
The III-IV transition shifts to higher
values of $F_D$ as $\tau$ increases, as shown by the shift in
the $d\langle V\rangle/dF_D$ peak 
in Fig.~\ref{fig:11}(b).
In phase IV, $P_{6}$ gradually decreases with increasing $\tau$ up to  
$\tau = 1 \times 10^4$, after which $P_6$ begins to increase again when
phase IV is replaced by phase V as shown previously.  
The transition to phase V is marked by a weak
local maximum in $d\langle V\rangle/dF_{D}$.

We can characterize the dynamics of  the dense phase
in terms of three
driving force regimes.
At small drives, $F_{D} < 1.25$, phases I and II appear.
For intermediate values, $ 1.25 < F_{D} < 5.5$,
the system 
is predominately in phase III.
At high drives of $F_{D} > 5.5$, phases IV and V occur.

\begin{figure}
\includegraphics[width=\columnwidth]{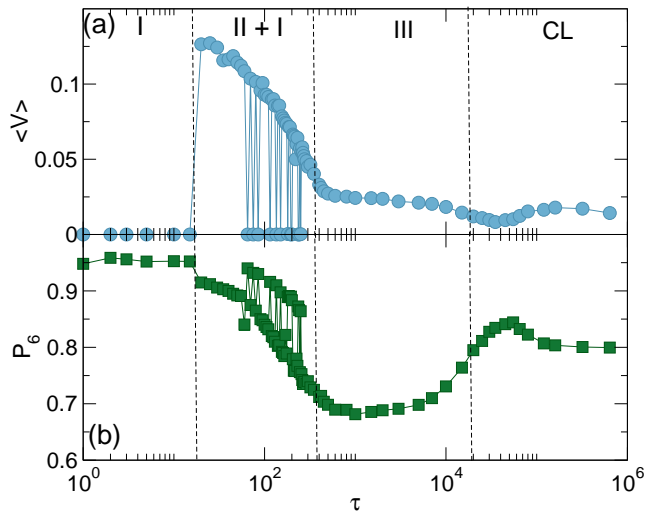}
\caption{ 
  (a) $\langle V\rangle$ vs $\tau$
  at $F_{D} = 0.15$ and $\phi = 0.848$.
  (b) The corresponding
  $P_{6}$ vs $\tau$.
  The system always reaches phase I
  for $\tau < 20$
  and phase II
  for $ 20 < \tau < 65$,
  while for $65 < \tau < 300$ the system can settle into either phase I or phase II.
  At larger $\tau$, phase III flow is stable, and for
  $\tau>1.5 \times 10^4$, phase CL flow occurs.
}
\label{fig:13ab}
\end{figure}

\begin{figure}
\includegraphics[width=\columnwidth]{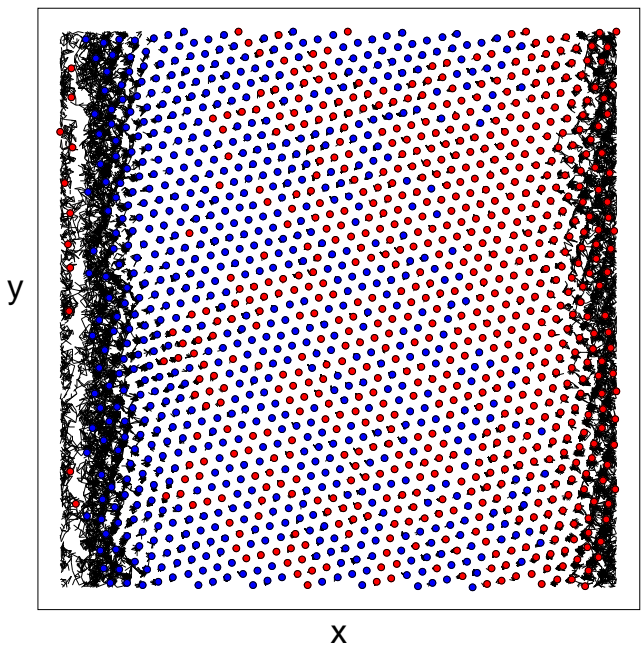}
  \caption{Trajectories over a fixed time period (lines) and instantaneous particle
    positions of the $\sigma_i=+1$ (blue circles) and $\sigma_i=-1$ (red circles)
    run-and-tumble
    particles
    for the system in Fig.~\ref{fig:13ab} at $\tau=200$ and $\phi=0.848$, which reaches
    a reentrant jammed phase I.
    For clarity, in this image we have reduced the radii of the circles representing
    the particles in order to make the trajectories visible.
Here there is a coexistence of a jammed state with a liquid.
}
\label{fig:14}
\end{figure}

In Fig.~\ref{fig:13ab}(a) we
plot $\langle V\rangle$ versus $\tau$ for a system with
$\phi  = 0.848$ and $F_{D} = 0.15$, and we show the corresponding
$P_6$ versus $F_D$ curve in Fig.~\ref{fig:13ab}(b).
For $\tau < 20$, the system always
forms a jammed phase I state
with $\langle V\rangle = 0.0$ and a high value of $P_{6}$.
Previous work with $\tau=0$ showed that phase II followed phase I upon
increasing $F_D$ \cite{19},
while in Fig.~\ref{fig:13ab} with fixed $F_D$,
phase II occurs for $20 < \tau < 60$ as indicated by the high
value of $\langle V\rangle$ in this regime.
We find that when $60 < \tau < 300$,
the system can organize into either the jammed phase I or the phase separated state
(phase II) as shown
by the jumps in $\langle V\rangle$ between $\langle V\rangle=0$ and
$\langle V\rangle \approx V_0$, the free flow velocity.
The plot in Fig.~\ref{fig:13ab} was obtained from individual realizations for each
value of $\tau$;
however,
if we average the value of $\langle V\rangle$ over many different realizations for
each $\tau$, we obtain $\langle \tilde V\rangle=0.05$
in this fluctuating regime since the system is in phase I for half of the realizations
and in phase II for the other half.
The reentrant behavior of phase I arises due to an effect similar to
freezing by heating \cite{46},
since the increase in the run time makes the particle act as if it had an
effectively larger radius, making the system susceptible to jamming.
In Fig.~\ref{fig:14} we show the particle positions and trajectories in 
the reentrant phase I
for $\tau = 200$.
Here the jammed phase takes the form of a triangular lattice, while in the
low density region, the particles are moving in a liquidlike fashion.
The appearance of the mixed phase I+phase II regime
is probably strongly size dependent, similar to the observation that
the freezing by heating
phenomenon
is enhanced by confinement \cite{46}.   
For $\tau>300$
in Fig.~\ref{fig:13ab},
the activity is large enough to break apart the crystalline structure
of phase I
and the system enters the disordered mixing
phase III, which coincides with a dip in $P_{6}$ and 
a drop in $\langle V\rangle$.
For $\tau > 1.5 \times 10^4$,
an increase in $P_{6}$
coincides with the onset of the activity-included
clustering phase CL
in which $\langle V\rangle$ remains low.
We find similar behavior as a function of $\tau$ over the range
$0 < F_{D} < 0.2$,
with the extent of the jammed phase I
increasing as $F_D$ decreases.
               
\begin{figure}
  \includegraphics[width=\columnwidth]{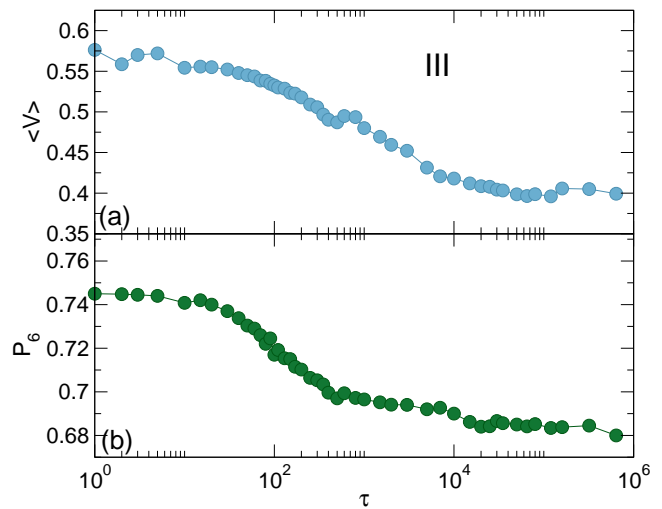}
  \caption{
    (a) $\langle V\rangle$ vs $\tau$ at $F_D=1.5$ and $\phi=0.848$ where the
    system is always in phase III.  (b) The corresponding $P_6$ vs $\tau$.
    }
  \label{fig:13cd}
\end{figure}

In Fig.~\ref{fig:13cd}(a,b)
we show $\langle V\rangle$ and $P_{6}$ versus $\tau$
for samples with $\phi = 0.848$ at $F_{D} = 1.5$,
where the system 
is always in phase III.
Here $\langle V\rangle$ gradually decreases from
$\langle V\rangle=0.575$ at $\tau = 1.0$ to
$\langle V\rangle=0.47$ with increasing 
$\tau$, while $P_{6}$ decreases from
$P_6=0.74$ to $P_6=0.69$.
We find similar behavior
for higher $F_{D}$ up to 
$F_{D} = 5.5$.

\begin{figure}
\includegraphics[width=\columnwidth]{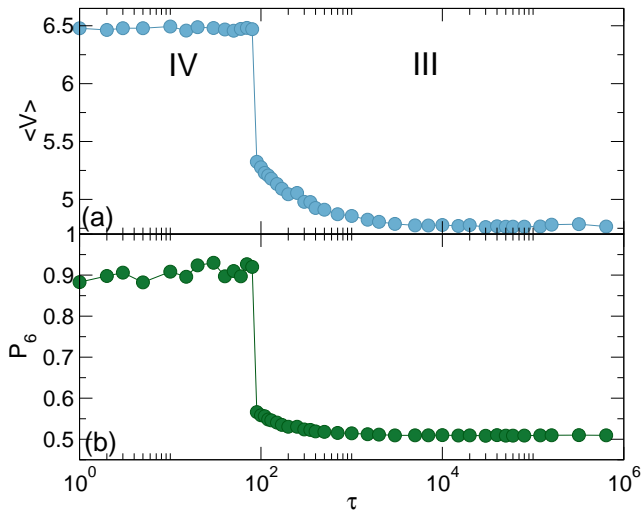}
\caption{
  (a) $\langle V\rangle$ vs $\tau$ at $F_{D} = 6.5$ and $\phi = 0.848$.
  (b) The corresponding
  $P_{6}$ vs $\tau$. Here the system undergoes a IV-III transition.
}
\label{fig:15ab}
\end{figure}

\begin{figure}
 \includegraphics[width=\columnwidth]{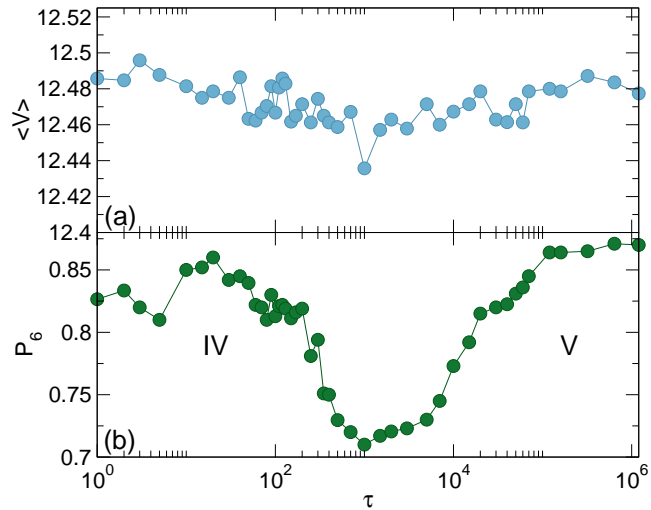}
 \caption{
   (a) $\langle V\rangle$ vs $\tau$ at $F_D=12.5$ and $\phi=0.848$.
   (b) The corresponding $P_6$ vs $\tau$.
   The IV-V transition appears as a dip in $P_6$, but there is
   is little change in $\langle V\rangle$ across the transition.
  }
  \label{fig:15cd}
\end{figure}

\begin{figure}
\includegraphics[width=\columnwidth]{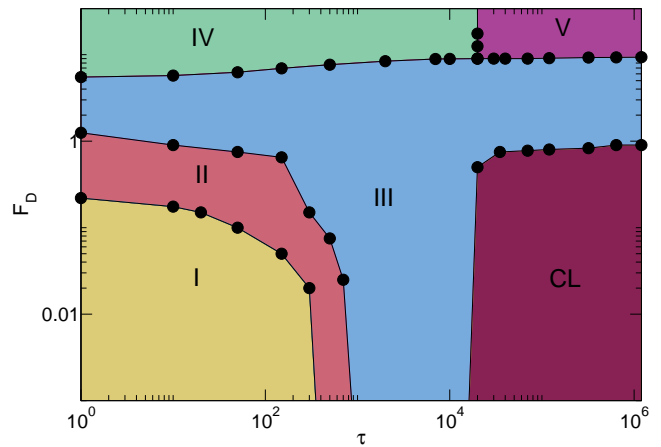}
\caption{
  Dynamic phase diagram as a function of
  $F_{D}$ vs $\tau$ at $\phi = 0.848$.
  In the region marked I the system always reaches the jammed phase I,
  while in the region marked II, the system sometimes reaches steady state
  phase II flow and sometimes enters a reentrant jammed phase I.
  Phase III is disordered mixing flow, phase IV is a laning state, phase V consists of
  laning cluster motion, and CL is the cluster phase.
}
\label{fig:16}
\end{figure}

In Fig.~\ref{fig:15ab} we plot $\langle V\rangle$ and $P_{6}$ versus
$\tau$ at $F_{D} = 6.5$ and $\phi = 0.848$ where the system is in the
laning phase IV up to $\tau = 100$.
Within phase IV, $\langle V\rangle \approx 6.5$
and $P_{6} \approx 0.9$.
At the transition to phase III,
there is an abrupt drop in both
$\langle V\rangle$ and $P_{6}$ when the onset of collisions between
the two species decreases the flow.
In Fig.~\ref{fig:15cd}(c,d) we show $\langle V\rangle$ and $P_6$ versus $\tau$
in the same system at
$F_{D} = 12.5$ where a IV-V transition occurs
near $\tau = 1.5 \times 10^4$.
At the transition,
a dip in $P_6$ appears
but there is little change in $\langle V\rangle$.
By conducting a series of simulations
and examining the features in
$P_{6}$ and $\langle V\rangle$ along with images of the
particle configurations,
we construct a dynamic
phase diagram as a function of $F_{D}$ versus $\tau$
for the $\phi = 0.848$ system,
as shown in Fig.~\ref{fig:16}.
In the region marked phase I, the system always reaches a jammed state,
while in the region marked phase II, the system is either in steady state
phase II flow or falls into a reentrant phase I jammed state.
Phases I and II appear only when
$\tau < 500$.
Phase IV occurs at large $F_{D}$ when $\tau < 15000$, phase 
CL occurs only when $\tau > 15000$,
and phase III separates phase II from phase IV, phase II from phase CL, and
phase CL from phase V.

\section{Summary} 
We have examined a two-dimensional binary system of particles
driven in opposite directions
where we introduce particle self-propulsion
in the form of run-and-tumble dynamics. 
Previous work on this system
in the non-active limit revealed four dynamic
phases: jammed, phase separated, disordered mixing flow, and laning flow.
At low particle densities, the non-active system
exhibits both laning and disordered flow phases.
As the activity is increased,
the laning phase transitions into a disordered
flow phase as indicated by both a drop
in the average mobility
and an increase in the frequency of particle-particle collisions.
The transition also appears
as a clear change in the velocity-force curve constructed using the average velocity
of one particle species as a function of the external drift force.
In terms of social systems,
such a transition can be compared to a change from
an orderly high mobility flow of agents such as pedestrians
to a low mobility panic state in which the agents collide.
At high drives we find a novel laning cluster state
in which the particles undergo both density phase segregation and
species phase segregation.
The laning cluster state remains stable well below the density at which an
activity-induced cluster state forms in an undriven system.
This suggests that applying external driving or shear can serve as
an alternative method of inducing cluster formation in an active system.
At high particle densities we find a total of six dynamic phases, 
including the jammed, phase separated, laning, and disordered flows,
the laning cluster state, and
an activity-induced cluster state
which appears for small external drift forces.
The activity can induce formation of a partially reentrant jammed state
at low drift forces
through a freezing by heating mechanism.
Our results show that binary driven active particles
exhibit a rich variety of behaviors. 
There are already several non-active experimental systems that 
can be modeled as binary driven systems,
and it may be possible to realize variations of active matter binary driven systems 
that would exhibit the behavior we describe.

\begin{acknowledgments}
We gratefully acknowledge the support of the U.S. Department of
Energy through the LANL/LDRD program for this work.
This work was carried out under the auspices of the 
NNSA of the 
U.S. DoE
at 
LANL
under Contract No.
DE-AC52-06NA25396 and through the LANL/LDRD program.
\end{acknowledgments}


\begin{thebibliography}{99}

\bibitem{1}
B. Schmittmann and R. K. P. Zia,
Driven diffusive systems. An introduction and recent developments,
Phys. Rep. {\bf 301}, 45 (1998).

\bibitem{2}
D. Helbing, I. J. Farkas, and T. Vicsek,
Freezing by heating in a driven mesoscopic system,
Phys. Rev. Lett. {\bf 84}, 1240 (2000).

\bibitem{3}
J. Dzubiella, G. P. Hoffmann, and H. L{\" o}wen,
Lane formation in colloidal mixtures driven by an external field,
Phys. Rev. E {\bf 65}, 021402 (2002). 

\bibitem{4}
R. R. Netz,
Conduction and diffusion in two-dimensional electrolytes,
Europhys. Lett. {\bf 63}, 616 (2003).

\bibitem{5}
J. Chakrabarti, J. Dzubiella, and H. L{\" o}wen,
Reentrance effect in the lane formation of driven colloids,
Phys. Rev. E {\bf 70}, 012401 (2004).

\bibitem{6}
T Glanz and H L{\" o}wen,
The nature of the laning transition in two dimensions,
J. Phys.: Condens. Matter {\bf 24}, 464114 (2012).

\bibitem{7}
K. Klymko, P. L. Geissler, and S. Whitelam,
Origin and macroscopic implications of lane formation in mixtures of
oppositely driven particles,
Phys. Rev. E {\bf 94}, 022608 (2016).

\bibitem{8}
A. Poncet, O. B{\' e}nichou, V. D{\' e}mery, and G. Oshanin,
Universal long ranged correlations in driven binary mixtures,
Phys. Rev. Lett. {\bf 118}, 118002 (2017).

\bibitem{9}
  M. E. Leunissen, C. G. Christova, A.-P. Hynninen, C. P. Royall, A. I. Campbell,
  A. Imhof, M. Dijkstra, R. van Roij and A. van Blaaderen,
Ionic colloidal crystals of oppositely charged particles,
Nature (London) {\bf 437}, 235 (2005).

\bibitem{10}
  T. Vissers, A. Wysocki, M. Rex, H. L{\" o}wen, C. P. Royall, A. Imhof and A. van Blaaderen,
Lane formation in driven mixtures of oppositely charged colloids,
Soft Matter {\bf 7}, 2352 (2011).

\bibitem{11}
T. Vissers, A. van Blaaderen, and A. Imhof,
Band formation in mixtures of oppositely charged colloids driven by an ac electric field,
Phys. Rev. Lett. {\bf 106}, 228303 (2011).

\bibitem{12}
K. R. S{\" u}tterlin, A. Wysocki, A. V. Ivlev, C. R{\" a}th, H. M. Thomas, M. Rubin-Zuzic,
W. J. Goedheer, V. E. Fortov, A. M. Lipaev, V. I. Molotkov, O. F. Petrov,
G. E. Morfill, and H. L{\" o}wen,
Dynamics of lane formation in driven binary complex plasmas,
Phys. Rev. Lett. {\bf 102}, 085003 (2009).

\bibitem{13}
C.-R. Du, K. R. S{\" u}tterlin, A. V. Ivlev, H. M. Thomas, and G. E. Morfill,
Model experiment for studying lane formation in binary complex plasmas,
EPL {\bf 99}, 45001 (2012).

\bibitem{14}
M. Moussaid, S. Garnier, G. Theraulaz, and D. Helbing,
Collective information processing and pattern formation in swarms, flocks, and crowds,
Topics in Cognitive Sci. {\bf 1}, 469 (2009).

\bibitem{15}
I. D. Couzin and N. R. Franks,
Self-organized lane formation and optimized traffic flow in army ants,
Proc. Roy. Soc. B {\bf 270}, 139 (2003).

\bibitem{16}
H. Ohta,
Lane formation in a lattice model for oppositely driven binary particles,
EPL {\bf 99}, 40006 (2012).

\bibitem{17}
T. Glanz, R. Wittkowski, and H. L{\" o}wen,
Symmetry breaking in clogging for oppositely driven particles,
Phys. Rev. E {\bf 94}, 052606 (2016).

\bibitem{18}
K. Ikeda and K. Kim,
Lane formation dynamics of oppositely self-driven binary particles: effects of
density and finite system size,
J. Phys. Soc. Jpn., {\bf 86}, 044004 (2017).

\bibitem{19}
C. Reichhardt and C. J. O. Reichhardt,
Velocity force curves, laning, and jamming for oppositely driven disk systems,
Soft Matter {\bf 14}, 490 (2018).

\bibitem{20}
C. Reichhardt and C. J. O. Reichhardt,
Cooperative behavior and pattern formation in mixtures of driven and nondriven colloidal assemblies,
Phys. Rev. E {\bf 74}, 011403 (2006).

\bibitem{21}
C. Reichhardt and C. J. O. Reichhardt,
Stripes, clusters, and nonequilibrium ordering for bidisperse colloids with repulsive interactions,
Phys. Rev. E {\bf 75}, 040402 (2007).

\bibitem{22}
A. Wysocki and H. L{\" o}wen,
Oscillatory driven colloidal binary mixtures: Axial segregation versus laning,
Phys. Rev. E {\bf 79}, 041408 (2009).

\bibitem{23}
M. Ikeda, H. Wada, and H. Hayakawa,
Instabilities and turbulence-like dynamics in an oppositely driven binary particle mixture,
EPL {\bf 99}, 68005 (2012).

\bibitem{24}
B. Heinze, U. Siems, and P. Nielaba,
Segregation of oppositely driven colloidal particles in hard-walled channels: A finite-size study,
Phys. Rev. E {\bf 92}, 012323 (2015).

\bibitem{25}
C. W. W{\" a}chtler, F. Kogler, and S. H. L. Klapp,
Lane formation in a driven attractive fluid,
Phys. Rev. E {\bf 94}, 052603 (2016).

\bibitem{26}
M. C. Cross and P. C. Hohenberg,
Pattern formation outside of equilibrium,
Rev. Mod. Phys. {\bf 65}, 851 (1993).

\bibitem{27}
T. Mullin,
Coarsening of self-organized clusters in binary mixtures of particles,
Phys. Rev. Lett. {\bf 84}, 4741 (2000).

\bibitem{28}
P. S{\' a}nchez, M. R. Swift, and P. J. King,
Stripe formation in granular mixtures due to the differential influence of drag,
Phys. Rev. Lett. {\bf 93}, 184302 (2004).

\bibitem{29}
C. Lozano, I. Zuriguel, A. Garcimart{\' i}n, and T. Mullin,
Granular segregation driven by particle interactions,
Phys. Rev. Lett. {\bf 114}, 178002 (2015).

\bibitem{30}
  M. C. Marchetti, J. F. Joanny, S. Ramaswamy, T. B. Liverpool, J. Prost, M. Rao, and
  R. A. Simha,
  Hydrodynamics of soft active matter,
  Rev. Mod. Phys. {\bf 85}, 1143 (2013).

\bibitem{31}
  C. Bechinger, R. Di Leonardo, H. L{\" o}wen, C. Reichhardt, G. Volpe, and G. Volpe,
  Active Brownian particles in complex and crowded environments,
  Rev. Mod. Phys. {\bf 88}, 045006 (2016).

\bibitem{32}
  Y. Fily and M. C. Marchetti,
  Athermal phase separation of self-propelled particles with no alignment,
  Phys. Rev. Lett. {\bf 108}, 235702 (2012).

\bibitem{33}
  G. S. Redner, M. F. Hagan, and A. Baskaran,
  Structure and dynamics of a phase-separating active colloidal fluid,
  Phys. Rev. Lett. {\bf 110}, 055701 (2013).

\bibitem{34}
  J. Palacci, S. Sacanna, A. P. Steinberg, D. J. Pine, and P. M. Chaikin,
  Living crystals of light-activated colloidal surfers,
  Science {\bf 339}, 936 (2013).

\bibitem{35}
  I. Buttinoni, J. Bialk{\' e}, F. K{\" u}mmel, H. L{\" o}wen, and C. Bechinger,
  Dynamical clustering and phase separation in suspensions of self-propelled
  colloidal particles,
  Phys. Rev. Lett. {\bf 110}, 238301 (2013).

\bibitem{36}
  M. E. Cates and J. Tailleur,
  When are active Brownian particles and run-and-tumble particles equivalent?
  Consequences for motility-induced phase separation,
  Europhys. Lett. {\bf 101}, 20010 (2013).

\bibitem{37}
  M. E. Cates and J. Tailleur,
  Motility-induced phase separation,
  Annu. Rev. Condens. Mat. Phys. {\bf 6}, 219 (2015).

\bibitem{38}
C. Reichhardt and C. J. O. Reichhardt,
Active microrheology in active matter systems: Mobility, intermittency, and avalanches,
Phys. Rev. E {\bf 91}, 032313 (2015).

\bibitem{39}
O. Chepizhko, E. G. Altmann, and F. Peruani,
Optimal noise maximizes collective motion in heterogeneous media,
Phys. Rev. Lett. {\bf 110}, 238101 (2013).

\bibitem{40}
O. Chepizhko and F. Peruani,
Diffusion, subdiffusion, and trapping of active particles in heterogeneous media,
Phys. Rev. Lett. {\bf 111}, 160604 (2013).

\bibitem{41}
C. Reichhardt and C.J. Olson Reichhardt,
Absorbing phase transitions and dynamic freezing in running active matter systems,
Soft Matter {\bf 10}, 7502 (2014).

\bibitem{42}
Cs. S{\' a}ndor, A. Lib{\' a}l, C. Reichhardt, and C. J. Olson Reichhardt,
Dynamic phases of active matter systems with quenched disorder,
Phys. Rev. E {\bf 95}, 032606 (2017).

\bibitem{43}
T. Bertrand, Y. Zhao, O. B{\' e}nichou, J. Tailleur, and R. Voituriez,
Optimized diffusion of run-and-tumble particles in crowded environments,
arXiv:1711.05209.

\bibitem{44}
  C. Reichhardt and C. J. O. Reichhardt,
  Active matter transport and jamming on disordered landscapes,
  Phys. Rev. E {\bf 90}, 012701 (2014).

\bibitem{45}
C.J. Olson Reichhardt and C. Reichhardt,
Avalanche dynamics for active matter in heterogeneous media,
New J. Phys. {\bf 20}, 025002 (2018).

\bibitem{46}
D. Helbing, I. J. Farkas, and T. Vicsek,
Freezing by heating in a driven mesoscopic system,
Phys. Rev. Lett. {\bf 8}4, 1240 (2000).

\bibitem{47}
N. Bain and D. Bartolo,
Critical mingling and universal correlations in model binary active liquids,
Nature Commun. {\bf 8}, 15969 (2017).

\end{thebibliography}
\end{document}